\documentclass[11pt,a4paper]{article}
\usepackage[hyperref]{acl2020}
\usepackage{times}
\usepackage{enumitem}
\usepackage{graphicx}
\usepackage{latexsym}
\usepackage{tabularx}

\usepackage{microtype}

\aclfinalcopy 

\title{Read it to me: An emotionally aware Speech Narration Application}

\author{Rishibha Bansal \\
  \texttt{bansalr@usc.edu}\\}

\date{}
\begin{document}
\maketitle
\begin{abstract}
    In this work we try to perform emotional style transfer on audios. In particular, MelGAN-VC architecture is explored for various emotion-pair transfers. The generated audio is then classified using an LSTM-based emotion classifier for audio. We find that "sad" audio is generated well as compared to "happy" or "anger" as people have similar expression of sadness.
\end{abstract}
\section{Introduction}
\begin{figure*}[!htp]
    \centering
    \includegraphics[width=\textwidth]{"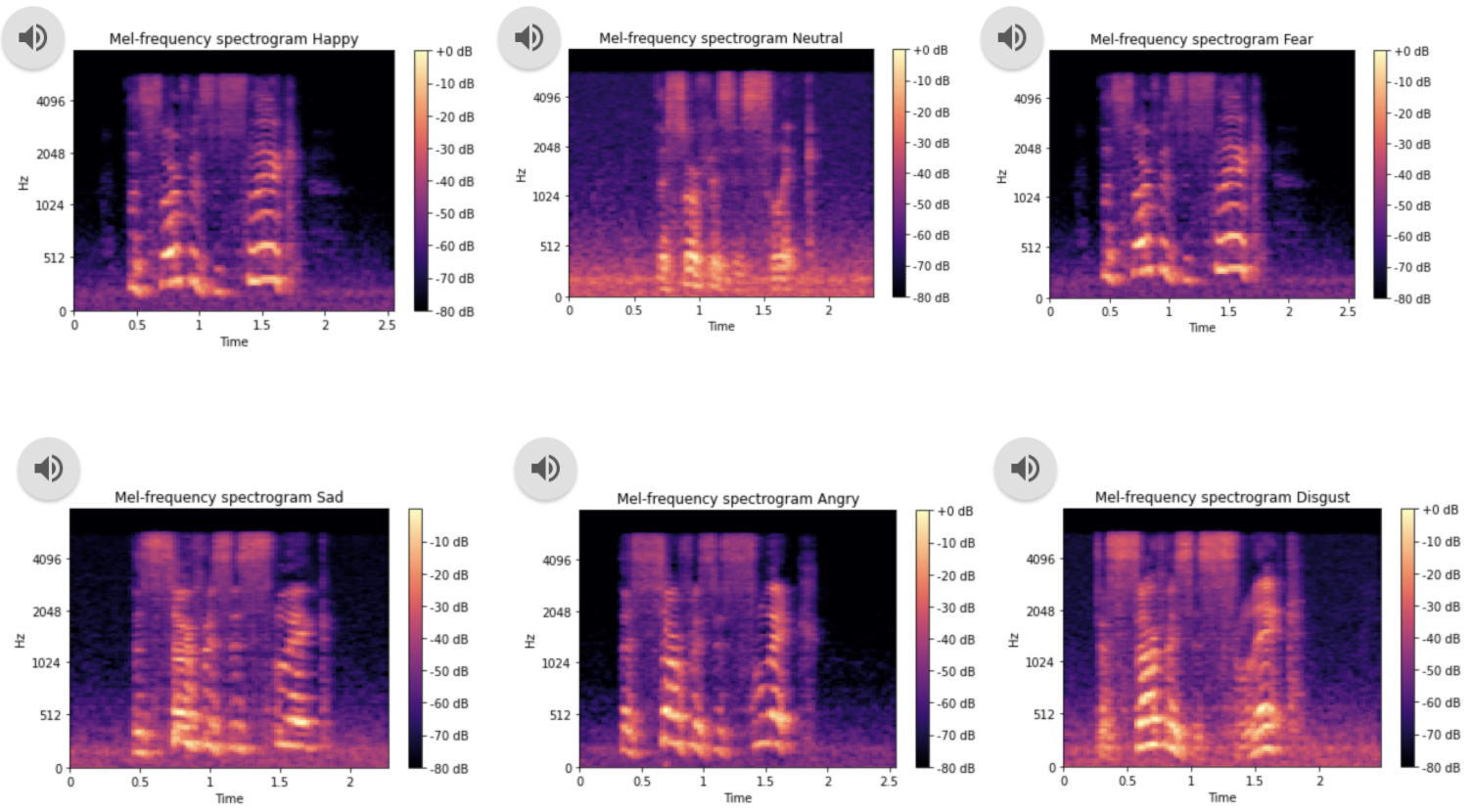"}
    \caption{Melspectrograms of 6 emotions by same actor saying "The surface is slick."}
    \label{fig:mel}
\end{figure*}

Most of the current Text to Speech(TTS) systems adopt a neutral tone for any text regardless of the content. Such neutral tones leave room to work on the humanization of the synthetic voices. One key factor could be adding emotions to the generated voice. Doing so may have potential benefits for auto-audio book creation as well as help people with speech impairments. In real-world applications, the emotional tone should be inferred from the text itself using sentiment analysis. However, the scope of this particular project is that, given a target emotion, text, and a reference speaker audio (it does not have to be for the exact text), generate speech for the given text in given emotional tones in the voice of the reference speaker. The modes involved in the project are text, speech, vision, and audio. As can be seen from Figure \ref{fig:mel}, the same person speaking the same sentence with different emotions will produce different spectrograms. It can be noticed that Mel spectrograms are informative with regards to emotions; for example, neutral tone utterances have low frequencies; angry and disgusted, which are excited emotions, have high frequencies. The architecture proposed in this work (Figure \ref{fig:arch}) is to perform style transfer (where style is emotions) using spectrograms.

\section{Literature Survey}
The project has two key components - 1. Generating speech from the text(TTS), 2. Emotional speech synthesis. For TTS, this project relies on the existing state-of-the-art models, the support for which has been integrated into Python. For emotional speech synthesis, existing architectures are studied and used as references. 
\subsection{Text to Speech(TTS)}
There are many existing types of research available for Text To Speech. One of them is made available by Google with multilingual support, called Tacotron 2 \cite{shen2018natural}, with several implementations, including one by Nvidea\footnote{https://github.com/NVIDIA/tacotron2}. This architecture consists of two modules: a recurrent seq2seq LSTM model with attention which predicts a Mel spectrogram from text and a WaveNet  \cite{oord2016wavenet} based vocoder to convert the Mel spectrogram back to time-domain waveforms.\\
Another paper of interest for TTS is Transfer learning from speaker verification to multispeaker text-to-speech synthesis  \cite{jia2018transfer} which highlights an architecture that is used to synthesize TTS with a given voice sample of just a few seconds. This powerful architecture includes three independently trained modules: one for a speaker encoder trained on thousands of speaker samples to generate a speaker embedding from just a few seconds of audio sample; two, a Tacotron 2 model conditioned on the speaker encodings to generate a Mel spectrogram and three, a WaveNet based vocoder to convert the Mel spectrogram into waveforms.
The TTS model used for this project incorporates this same architecture using Coqui AI\footnote{https://github.com/coqui-ai/TTS}.
\subsection{Emotional Speech Synthesis}
For emotional speech synthesis, research in emotional speech generation and style transfer, in general, is studied. Controllable emotion transfer for end-to-end speech synthesis \cite{li2021controllable} is used for emotional speech synthesis in Mandarin. The research involves modifying the Tacotron 2 network to condition the encoder on the emotional embedding of reference audio and then minimizing the loss between the emotional embedding of generated audio and reference audio. The data used to train the model includes 14 hours of Mandarin speech samples by a Chinese actress. The results appear promising.\\
Another paper, MelGan \cite{pasini2019melgan} deals with style transfer in audio using a generator, discriminator, and a Siamese network to allow transitions between substantially different domains of audio and audio samples of variable length. \\
A revolutionary work, Image style transfer using convolutional neural networks \cite{gatys2016image} deals with style transfer in images, adopting a style of an artist to regular images. The authors propose a model that balances two kinds of losses - content loss and style loss. The model is CNN-based. \\
Similar ideology can be applied to audios, and this project draws concepts from this paper.

\section{Datasets}
\begin{figure*}[!htp]
    \centering
    \includegraphics[width=\textwidth]{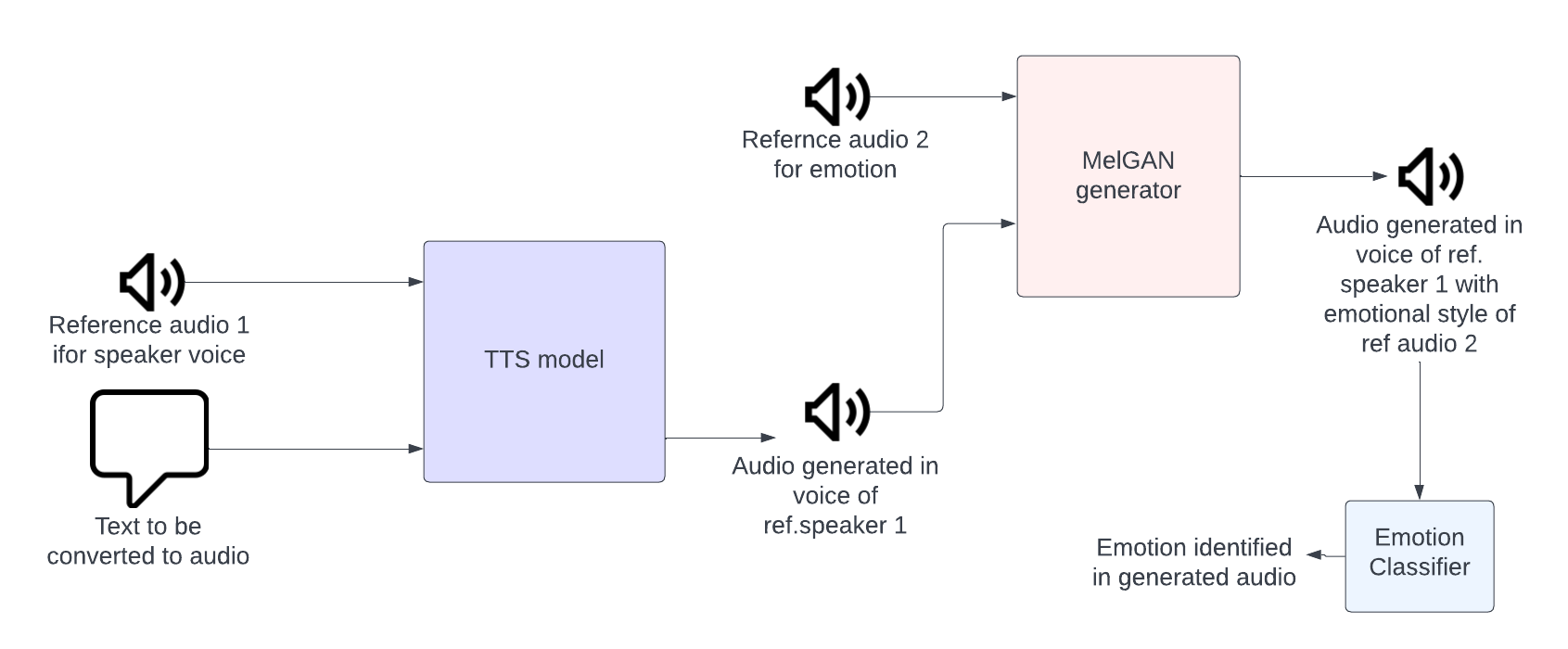}
    \caption{Proposed architecture}
    \label{fig:arch}
\end{figure*}
For this project, four datasets consisting of emotion-based audios have been curated. The total number of audio samples is 12,162 of comparable lengths.
\begin{enumerate}[noitemsep]
\item CREMA-D \citep{cao2014crema} dataset consists of 7,442 original clips of a few seconds each from 91 actors. The actors utter sentences from a collection of 12 sentences in six emotions (Anger, Disgust, Fear, Happy, Neutral, and Sad). The dataset has a mix of male and female actors aged 20-74 and of varying ethnicities. The audio quality is mixed. The emotions have different intensity levels, but for this project, all levels are treated the same
\item RAVDESS \cite{livingstone2018ryerson} dataset consists of 1440 samples by 12 male and 12 female actors reading from a collection of 2 sentences. In addition to the six emotions mentioned in CREMA-D, it also includes the "surprise" emotion.
\item SAVEE \cite{jackson2014surrey} dataset contains 480 audio clips by four adult males. It includes all seven emotions (Anger, Disgust, Fear, Happy, Neutral, Neutral, and Sad).
\item TESS \cite{dupuis2010toronto} dataset contains 2800 clips by two female actors and again includes all seven emotions. The sentences are in the form of "Say the word \_" from a pool of 200 words.
\end{enumerate}
The imbalance created by the SAVEE dataset being all-male is balanced out by the TESS dataset being all female and the CREMA-D dataset (43 females, 48 males). All audio clips are 1-3 seconds long in each dataset, creating symmetry. Out of all seven emotions, all emotions have around 2k samples each except surprised (592 records). The "surprise" emotion is ignored to handle this imbalance. 

\section{Method}\label{section:method}

Although the approach mentioned in Controllable emotion transfer for end-to-end speech synthesis \cite{li2021controllable} is inspired and produced good results in Mandarin, the current resource constraints prevent this project from replicating the architecture for English.
Two approaches are considered for this work. First is Neural style transfer. Another is Style transfer using MelGAN-VC \cite{pasini2019melgan}. The main focus is on MelGAN style transfer. The end to end pipeline includes taking text and reference audio\textsubscript{1} (for speaker's voice) as input to a TTS model; taking the output of TTS and another reference audio\textsubscript{2} (for emotional style) as input for the MelGAN generator; finally, classifying if the generated audio has same emotion as intended by reference audio\textsubscript{2}. The proposed architecture can be seen in Figure \ref{fig:arch}.

\subsection{Neural style transfer for emotions}
The approach is to apply audio style transfer similar to Image style transfer \cite{gatys2016image} after TTS has been performed on the given text using the given reference audio using pre-trained models by Coqui AI (as described in Section \ref{section:res}). The architecture involves single-layer CNN, which tries to reduce the style loss using Gram matrices between the two spectrograms-reference audio and Coqui-generated audio. The model does not perform that well. Detailed analysis is found in section \ref{section:res}. Some samples generated by this model can be found here\footnote{https://docs.google.com/presentation/d/1F74DTFldMDld
iuctVkgU31tLAIMbCdBhvmW3cQaQvTA/edit?usp=sharing}.

\subsection{Style transfer using MelGAN-VC}
\begin{figure*}[!htp]
    \centering
    \includegraphics[width=\textwidth]{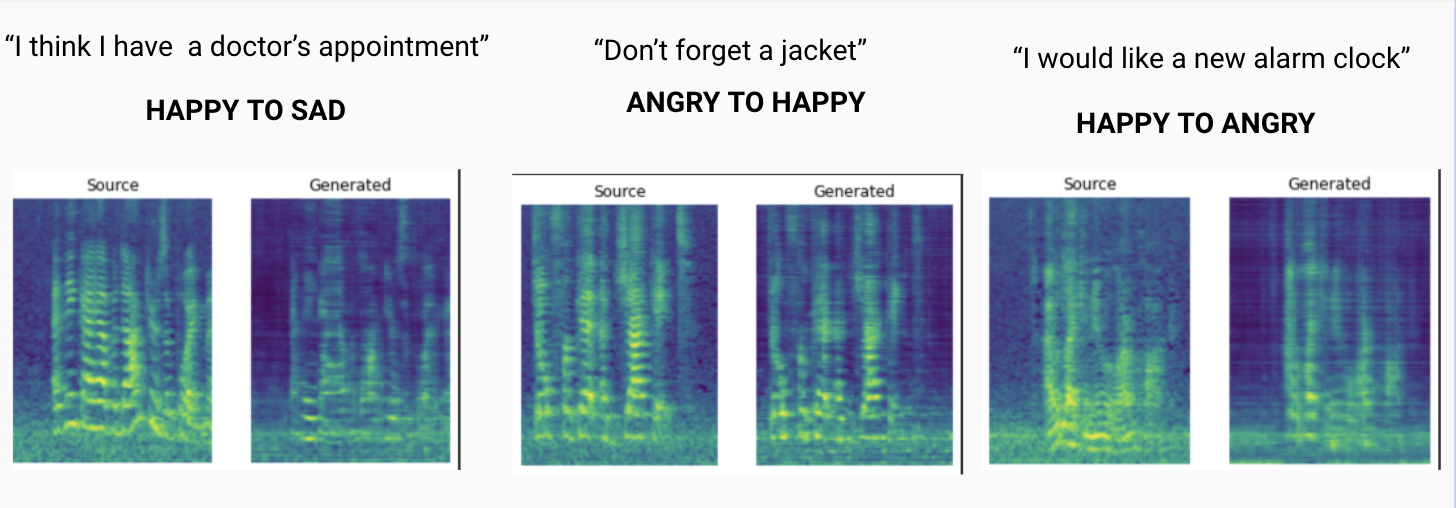}
    \caption{Some Melspectrograms generated by MelGANs where source represents original audio and Generated represents audio file with target emotion with the same linguistic content as original audio}
    \label{fig:final_mel}
\end{figure*}

This model consists of a Generator and a Discriminator like all GANS. However, in addition to these, a siamese network is used to maintain linguistic information during the conversion by the generator. The model takes the spectrograms of audio generated using Coqui for the required text (audio 1) and converts them to the spectrogram of reference audio style, keeping the content the same. The model learns the distribution of target emotion in our case by training on thousands of samples of target emotion. The model minimizes three losses - generative, discriminative, and siamese. The network has been borrowed from the original work MelGAN-VC \cite{pasini2019melgan} which includes 3 U-net architectures. For this work, six different MelGANs were trained for different target emotions. Detailed analysis is found in section \ref{section:res}. Some samples generated by this model can be found here\footnote{https://docs.google.com/presentation/d/1oUaDJkVSyw8
CshVO3eilSkPQmx1gqtvra58K12YLyO4/edit?usp=sharing}. A few samples of spectrograms generated by MelGANs can be seen in Figure \ref{fig:final_mel}.  

\subsection{Emotion Classifier}
To test the efficiency of MelGAN-generated audios, an emotional classifier is built - trained on six emotions (happy, sad, angry, fearful, disgusted, and neutral). The classifier has a 128-unit input layer and two 256-unit FC layers with ReLu activation. The final softmax layer outputs the probability of the six emotion types. The classifier is used to evaluate finally generated outputs by MelGAN by comparing classified emotion to intended emotion. The classifier is trained for ten epochs on the entire dataset (except TESS), where MFCC features are extracted from the audio files with window size 25ms, hop length 10ms, and the number of Melspectrogram filters as 40. Adam optimizer is used with a learning rate of 0.0001. The model gives an accuracy of 49.6\% and an F1 score of 48.3\%.      

\section{Experiments}\label{section:res}

\begin{table*}
\begin{center}
\begin{tabular}{ ||c|c|c|| }
\hline \textbf{Model} & \textbf{Accuracy} & \textbf{F1} \\ \hline
 All emotions(original audio) classifier & 49.6 & 48.3\\ 
 MelGAN Happy to Sad & 43.8 & 10.3 \\ 
 MelGAN Happy to Angry & 46.7 & 12.7\\ 
 MelGAN Angry to Sad & 32.7 & 8.22 \\
 MelGAN Angry to Happy & 16.95 & 4.83 \\
 MelGAN Sad to Happy & 33.33 & 8.33 \\
 MelGAN Sad to Angry & 17.64 & 4.9 \\
 \hline
\end{tabular}
\end{center}
\caption{\label{comparetable} Comparison of different MelGANs by testing generated audio using emotional classifier with target emotion as test labels.}
\label{table:accuracy}
\end{table*}

\subsection{Audio generation from text}
A Python module by Coqui AI has been used for the TTS module. This toolkit includes all state-of-the-art models for the three sub-modules (Text2Spec, Speaker Encoder, Vocoder). After testing out various combinations, for this project, Tactotron 2 \cite{shen2018natural} has been chosen for the Text2Spec module with Wavegrad \cite{chen2020wavegrad} as the Vocoder model. When fed with reference audio and text, the TTS model generates text in the voice of reference audio in neutral tones.

\subsection{Feature Extraction}
The audio clips have been subjected to various feature extractions like amplitude envelop, zero crossing rates, and Mel frequency cepstral coefficients (MFCCs). For the initial iteration, MFCCs are used for emotion embedding extraction. Later on, other features or a combination of them could be tried. MFCCs are extracted with a 25ms window length and a 10-second hop with 40 Mel filters. These hyperparameters are adjusted according to fine-tuning based on the performance of the emotion classifier. A brief analysis of Melspectrograms shows that audio for the same text by the same speaker is affected by emotions, as can be seen in figure \ref{fig:mel}. It can be noticed that Mel spectrograms are informative with regards to emotions; for example, neutral tone utterances have low frequencies, and angry and disgusted, which are excited emotions, have high frequencies.
\subsection{MelGAN training}
For this work, only three emotions - Happy, Angry, and Sad are generated using MelGANs. A total of six MelGANs have been trained on the following pairs of emotion styles:
\begin{enumerate}
\item MelGAN learned to produce \textbf{sad} emotion using 1100 audio samples of Happy emotions (Crema-D) as source emotion and 1100 audio samples of sad emotions (Crema-D) as training. 
\vspace{-8px}
\item MelGAN learned to produce \textbf{angry} emotion using 1100 audio samples of Happy emotions (Crema-D) as source emotion and 1100 audio samples of angry emotions (Crema-D) as training.
\vspace{-8px}
\item MelGAN learned to produce \textbf{sad} emotion using 1100 audio samples of angry emotions (Crema-D) as source emotion and 1100 audio samples of sad emotions (Crema-D) as training.
\vspace{-8px}
\item MelGAN learned to produce \textbf{happy} emotion using 1100 audio samples of sad emotions (Crema-D) as source emotion and 1100 audio samples of happy emotions (Crema-D) as training.
\vspace{-8px}
\item MelGAN learned to produce \textbf{angry} emotion using 1100 audio samples of sad emotions (Crema-D) as source emotion and 1100 audio samples of angry emotions (Crema-D) as training.
\vspace{-8px}
\item MelGAN learned to produce \textbf{happy} emotion using 1100 audio samples of angry emotions (Crema-D) as source emotion and 1100 audio samples of happy emotions (Crema-D) as training.
\end{enumerate}

Each model is trained for 20 epochs using 0.0002 as the learning rate, 3 generator updates per discriminator updates, and a batch size of 16. Each model is tested against 177 samples of source emotion. The 177 target emotion files generated by each model are sent to the classifier for evaluation to understand the quality of emotions produced by the MelGAN. A comparative study is found in table \ref{table:accuracy}. For example, a MelGAN trained on happy to sad should ideally produce all sad sounding audios. However, the classifier identified the generated audio as sad only 43.8\% of the time and the remaining times as one of the other five emotions.
 
\section{Discussions}

For \textbf{Neural style transfer of emotions}, it was not apparent what the model considered "style"; the pitch and tone or the emotion. Hence, two separate scenarios were considered  -  style transfer between audios corresponding to the same speaker and style transfer between audios corresponding to different speakers. Neither of the techniques performed well. The target linguistic content was the same as the source, but the reference emotional style was not mapped well in either case.\\
For \textbf{Style transfer using MelGAN-VC}, as we can see from table \ref{table:accuracy}, sad and angry emotion audios were generated with higher quality than happy emotion audios by MelGANs. The reason anger performed well could be due to the coarseness introduced in the audio by MelGAN. More epochs may help in removing this issue. In particular, the mapping from happy to sad and happy to angry emotion performed well (subjectively speaking). Sadness also performed well when the style was transferred from anger audios. This shows there is a distinctive distribution of sad emotion that the model was able to learn, despite there being 91 different speakers. Our results also show that people tend to display sadness in a similar way. The same cannot be said for the happy emotion, as not everyone expresses happiness in a similar way. In this case, the 91 different speakers and their unique way of expressing happiness prevented the model from learning a good distribution of happiness and anger emotion. It is also to be noted that training on the same speaker with multiple samples pertaining to a particular emotion would have been more beneficial.\\
Another task was to determine if punctuation matters in the text. Preliminary testing shows that punctuation does not affect emotions for existing TTS models.

\section{Conclusion}

This work explores two approaches for generating audio for a text sample in a particular emotional style. MelGAN-VC outperformed Neural style transfer for emotional style transfer. MelGAN-VC trained on audio samples from 91 speakers performed exceptionally well for the "sad" emotion because different people tend to express sadness in the same way. The same is not true for happiness and anger emotion. Training on the same speaker with multiple samples pertaining to an emotion would have been more beneficial. In this work, only three emotions were considered- happiness, anger, and sadness. We plan to extend this work to other emotions in the future.

\bibliography{anthology,acl2020,references}
\bibliographystyle{acl_natbib}

\end{document}